\documentclass[a4paper, 11pt]{article}

\usepackage[utf8]{inputenc} % allow utf-8 input
\usepackage[T1]{fontenc}    % use 8-bit T1 fonts
\usepackage{hyperref}       % hyperlinks
\usepackage{url}            % simple URL typesetting
\usepackage{booktabs}       % professional-quality tables
\usepackage{nicefrac}       % compact symbols for 1/2, etc.
\usepackage{lipsum}
\usepackage{microtype}      % microtypography (loaded after other packages)
\usepackage{graphicx}
\usepackage{float}
\usepackage{placeins}
\usepackage{amsmath}
\usepackage{amssymb}
\usepackage[normalem]{ulem}
\usepackage{lineno}
\usepackage{longtable}
\usepackage[style=numeric-comp,sorting=none]{biblatex}
\usepackage{multirow}
\usepackage{authblk}
\usepackage{lineno}
\usepackage{color,soul}
\usepackage[colorinlistoftodos, textsize=small]{todonotes}
\newcommand{\beginsupplement}{%
  \setcounter{table}{0}
  \renewcommand{\thetable}{S\arabic{table}}%
  \setcounter{figure}{0}
  \renewcommand{\thefigure}{S\arabic{figure}}%
  \setcounter{equation}{0}
  \renewcommand{\theequation}{S\arabic{equation}}%
}

\title{Care Trajectories Are Linked to Mental Health and Mortality in Cancer Patients}
\author[1,2]{Simon D. Lindner}
\author[3,4,5,6]{Elisabeth L. Zeilinger}
\author[3,5,6]{Amelie Fuchs}
\author[3]{Simone Lubowitzki}
\author[1,2,7,8*]{Peter Klimek}
\author[3,9,10]{Alexander Gaiger}

\affil[1]{Section for the Science of Complex Systems, CeMSIIS, Medical University of Vienna, Vienna, Austria}
\affil[2]{Complexity Science Hub Vienna, Vienna, Austria}
\affil[3]{Division of Hematology and Hemostaseology, Department of Medicine I, Medical University of Vienna, Vienna, Austria}
\affil[4]{Division of Health Psychology, Faculty of Psychology, Karl Landsteiner University of Health Sciences, Krems, Austria}
\affil[5]{Academy for Ageing Research, Haus der Barmherzigkeit, Vienna, Austria}
\affil[6]{Department of Clinical and Health Psychology, Faculty of Psychology, University of Vienna, Vienna, Austria}
\affil[7]{Supply Chain Intelligence Institute Austria, Vienna, Austria}
\affil[8]{Division of Insurance Medicine, Karolinska Institutet, Department of Clinical Neuroscience, Karolinska}
\affil[9]{Comprehensive Cancer Center, Medical University of Vienna, Vienna, Austria}
\affil[10]{Ludwig Boltzmann Institute for Hematology and Oncology, Medical University of Vienna, Vienna, Austria}

\affil[*]{Corresponding author: peter.klimek@meduniwien.ac.at}

\date{}

\begin{document}

\maketitle
\begin{abstract}
Treatment of haematologic malignancies often involves long and complex care pathways that vary substantially across patients. How baseline mental health characteristics relate to these heterogeneous trajectories, and whether trajectory patterns carry prognostic information beyond established clinical and socioeconomic factors, remains poorly understood. In this study, we introduce a  fully interpretable time-analysis framework that leverages these temporal dynamics to enhance our understanding of outcomes of patients treated for haematologic malignancies. We analyzed longitudinal care patterns observed in up to 37 years from over 8,000 patients using Dynamic Time Warping (DTW) and Hierarchical Clustering. This resulted in the identification of nine distinct, non-trivial patient clusters encoding robust healthcare utilization trajectories.

We evaluated the prognostic utility of these clusters by including them, alongside conventional covariates (socioeconomic variables, age, sex, and diagnoses), as features in generalized linear models for mortality prediction and for assessing associations with baseline anxiety and depression scores. The inclusion of these clusters significantly enhanced the model's performance and maintained significance even after adjusting for classic covariates.

Notably, all eight trajectory clusters showed significantly higher odds of mortality compared to the reference group, which is characterized by short observation periods (median 6.3 years) and the lowest mortality rate (31.5\%). These high-mortality clusters fall into two distinct patterns: clusters with complex, long-term care pathways comprising up to 196 health-related events (mortality OR up to 3.38, 95\% CI 2.13--5.37), and clusters with shorter but severe trajectories (median 78 events; mortality OR 2.32, 95\% CI 1.82--2.97). Notably, the two complex-care clusters also showed significantly lower baseline anxiety scores, revealing a divergent relationship between trajectory complexity, mortality risk, and baseline mental health. These findings demonstrate that incorporating temporal patient trajectories offers significant explanatory power for refining risk assessment in oncology.
\end{abstract}

\section{Introduction}
Cancer remains one of the leading causes of morbidity and mortality worldwide, placing a significant burden on patients and healthcare systems. Although advances in diagnostics and therapeutics have improved outcomes for many cancer types, survival rates and quality of life remain highly variable across patient populations \cite{bedirSocioeconomicDisparities2022, chenTrends5year2024, dixitHealthrelatedQuality2024, jiangGlobalPattern2021, ulibarri-ochoaDifferencesQuality2023}. A key challenge in oncology is effectively stratifying patients, i.e. identifying subgroups that differ in terms of prognosis, treatment requirements and risk of adverse outcomes. Traditionally, stratification has relied heavily on tumour biology, stage at diagnosis and demographic variables such as age and sex \cite{meyerStratificationSystem2025, riviereProstateSpecificAntigen2024, vanstigtIntervalCancer2025, winderAgeStratification2024}.
While these factors provide important information, they often fail to capture the dynamic and multifaceted nature of cancer care. In particular, little attention has been given to how the sequence and timing of visits, diagnoses, and treatments in patients' longitudinal care trajectories shape disease progression and clinical outcomes despite evidence of persistent disparities even in controlled clinical trial settings 
\cite{ungerPersistentDisparity2021}. Moreover, the complexity and duration of cancer care pathways may also relate to patients' psychological well-being, yet how different trajectory patterns associate with mental health characteristics remains largely unexplored.

Despite increasing recognition of the importance of temporal dynamics in healthcare data, their systematic integration into the identification of trajectory-based cancer phenotypes that capture both disease severity and mental health dimensions remains limited \cite{houCardiacRisk2021, leeDevelopingMachine2022, masonPredictionMetastatic2021, morinArtificialIntelligence2021, paikCondensedTrajectory2021, placidoDeepLearning2023, sunVMRAMaRAsymmetryAware2026}. Existing studies have typically focused on single cancer types, such as prostate or bladder cancer \cite{leeDevelopingMachine2022, masonPredictionMetastatic2021}, or on specific adverse outcomes, for example cardiac complications in oncology patients \cite{houCardiacRisk2021}. Broader investigations have characterized temporal disease trajectories across large patient populations \cite{hjaltelinVisualisingDisease2023, paikCondensedTrajectory2021} or used longitudinal data primarily for cancer risk prediction rather than patient stratification \cite{placidoDeepLearning2023, sunVMRAMaRAsymmetryAware2026}. Although some efforts have combined longitudinal real-world data across multiple cancer types \cite{morinArtificialIntelligence2021}, these approaches have not explicitly modeled temporal evolution for dynamic subgrouping. Similarly, although socioeconomic and psychosocial factors are well-established independent predictors of cancer outcomes \cite{gaigerCancerSurvival2022, ungerPersistentDisparity2021}, they are rarely incorporated into trajectory-based analyses. Consequently, while temporal information is increasingly being incorporated into predictive modeling, most stratification frameworks still rely on static representations of clinical or demographic variables, overlooking the sequential and evolving nature of cancer care pathways \cite{khaliqRefiningColorectal2022, liuRiskStratification2021, meyerStratificationSystem2025}. Closing these gaps is crucial for developing patient risk models that account for the temporal organization of care and its potential associations with both mortality and mental health.

Beyond static clinical indicators, understanding cancer outcomes requires consideration of the temporal dimension of care \cite{franssenAssociationTreatment2021, kouTreatmentIntervals2023, weiImpactDiagnosistoTreatment2025}. The sequence and timing of medical events, such as hospital admissions, outpatient visits, and treatment intervals, reflect both the biological progression of the disease and how care is organised and delivered. Variations in these trajectories may indicate differences in disease aggressiveness, treatment response, or access to healthcare services. For example, frequent or prolonged hospital visits may suggest a more advanced stage of the disease or complications, while irregular follow-up patterns may indicate problems with continuity of care. Capturing these longitudinal patterns provides a more nuanced understanding of patient heterogeneity, enabling the identification of dynamic phenotypes that combine severity with mental health characteristics and go beyond conventional demographic or tumour-based factors \cite{giannoulaIdentifyingTemporal2018, giannoulaExploringLongterm2024}. Incorporating temporal dynamics into predictive modelling can help to identify hidden structures in healthcare utilisation that may have important prognostic and therapeutic implications.

In addition to biological and clinical characteristics, socio-economic and psychosocial factors play a decisive role in shaping cancer outcomes \cite{faySocioeconomicInequalities2024, gaigerCancerSurvival2022, lawlerArealevelSocioeconomic2025, marocoRaceEthnicity2025, wangSocialSupport2024, zhaoSideEffects2022}. Socioeconomic status (SES) has been consistently linked to disparities in cancer incidence, access to timely care, and survival rates \cite{gaigerCancerSurvival2022, jiAssociationSocioeconomic2020, redondo-sanchezSocioEconomicInequalities2022, zeilingerImpactCOVID192022}. Patients from socioeconomically disadvantaged backgrounds often have a worse prognosis and poorer mental health, which highlights the pervasive impact of these determinants \cite{ungerPersistentDisparity2021}. Psychosocial factors such as depression, anxiety and post-traumatic stress symptoms can also influence treatment adherence, recovery and overall well-being \cite{fangSelfratedHealth2024, jeongEffectDepression2025, sengerRoleCoping2024, songMediatingRole2025}. Despite their recognised importance, these dimensions are often underrepresented in predictive models of cancer care. Combining socioeconomic and psychosocial variables with clinical and temporal data could reveal more comprehensive risk profiles, enabling interventions that are both medically and socially responsive.

Recent advances in data science and artificial intelligence have provided powerful tools for analysing complex healthcare data. Among these, dynamic time warping (DTW) has emerged as an effective method for comparing and clustering longitudinal patient trajectories. DTW was originally developed for signal processing \cite{sakoeDynamicProgramming1978}, but has since been successfully adapted for use in the biomedical domain, where it enables the alignment and comparison of irregular and heterogeneous clinical sequences \cite{giannoulaIdentifyingTemporal2018, giannoulaExploringLongterm2024}. Studies applying DTW to patient records have demonstrated its ability to reveal temporal patterns in disease progression and healthcare utilisation, providing new insights into patient stratification \cite{forrestDynamicTime2025, fuIdentifyingCommon2025, giannoulaExploringLongterm2024}. Clustering approaches based on DTW enhance this capability further by grouping patients with similar medical trajectories, thereby identifying subpopulations that are not apparent through conventional analyses. Although promising results have been reported in areas such as chronic disease management and breast cancer survivorship \cite{giannoulaExploringLongterm2024}, most previous applications have focused on specific conditions or limited datasets, rarely integrating socioeconomic or psychosocial dimensions into the modelling framework.

This study aims to address these gaps by introducing a framework that combines temporal visit patterns with socio-economic and psychosocial factors. Using a substantial dataset comprising over 8,000 patients over several decades, we employ Dynamic Time Warping and hierarchical clustering to identify distinct groups based on healthcare utilisation trajectories. These trajectory-based clusters are then incorporated into predictive models alongside conventional covariates, such as demographic and diagnostic variables. By incorporating longitudinal care patterns alongside broader contextual factors, our approach seeks to identify trajectory phenotypes with distinct mortality risk and mental health profiles, advancing beyond conventional stratification approaches.
The specific aim of this study is therefore to develop and evaluate a trajectory-based framework that integrates temporal visit patterns with socioeconomic and psychosocial factors to identify clinically meaningful phenotypes among cancer patients.

\section{Methods and Data}

\subsection{Cohort and Data Description}
The study cohort comprised 8,120 distinct adult patients with a wide range of solid tumours and haematologic malignancies, recruited from the Clinic of Oncology and Hemostaseology at the General Hospital of Vienna, including both inpatients and outpatients across diverse disease stages and treatment modalities. Eligible participants were able to provide informed consent and complete study assessments. Data collection combined standardized self-report instruments and medical record abstraction.

An overview of the cohort characteristics, including clinical status, tumour types, and demographics, is provided in Table~\ref{tab:merged_characteristics}. The dataset spans from 1985 to 2022. For each patient, the comprehensive dataset includes details on tumor type, age, visit history, visit diagnoses, and responses to standardized self-report instruments, including the Hospital Anxiety and Depression Scale (HADS) for anxiety and depression, and the Post-Traumatic Symptom Scale (PTSS-10; also used in~\cite{zeilingerImpactCOVID192022}), a 10-item self-report instrument to assess post-traumatic stress symptoms, where each item is rated from zero to three (sum score 0--30; higher scores indicating higher symptom burden), alongside socioeconomic indicators. 

\begin{table}[H]
\centering
\caption{Cohort Characteristics: Clinical Status, Tumor Types, and Demographics}
\label{tab:merged_characteristics}
\small % Reduces font size slightly to fit vertically
\setlength{\tabcolsep}{10pt} % Adds nice spacing between columns

\begin{tabular}{lccc}
\hline
\textbf{Category} & \textbf{Male} & \textbf{Female} & \textbf{Overall} \\
\hline
\multicolumn{4}{l}{\textbf{Survival} \textit{(N=6522)}} \\
Survived & 50.0\% & 58.3\% & 54.7\% \\
\hline
\multicolumn{4}{l}{\textbf{Tumor Type} \textit{(N=6522)}} \\
Haematological & 29.0\% & 23.1\% & 25.8\% \\
Breast & 1.0\% & 32.7\% & 18.2\% \\
Lung & 14.9\% & 10.5\% & 12.5\% \\
Head and neck & 7.9\% & 3.2\% & 5.4\% \\
Colon/rectum & 7.3\% & 3.7\% & 5.3\% \\
Soft tissue & 6.1\% & 4.5\% & 5.2\% \\
Pancreas & 5.5\% & 4.8\% & 5.1\% \\
Brain & 5.1\% & 4.7\% & 4.9\% \\
Female genital organs & 0.0\% & 6.1\% & 3.4\% \\
Kidney/urinary tract & 5.0\% & 1.8\% & 3.2\% \\
Stomach/oesophagus & 4.3\% & 1.4\% & 2.7\% \\
Testis & 3.9\% & 0.0\% & 1.8\% \\
Prostate & 3.6\% & 0.0\% & 1.6\% \\
Thyroid & 1.6\% & 1.3\% & 1.4\% \\
Malignant melanoma & 1.8\% & 1.0\% & 1.4\% \\
Hepatobiliary & 1.6\% & 0.5\% & 1.0\% \\
Other & 1.4\% & 0.6\% & 1.0\% \\
\hline
\multicolumn{4}{l}{\textbf{Age Category} \textit{(N=6496)}} \\
$<$30 & 5.9\% & 3.8\% & 4.8\% \\
30-39 & 6.9\% & 8.8\% & 8.0\% \\
40-49 & 11.3\% & 15.8\% & 13.8\% \\
50-60 & 21.1\% & 22.9\% & 22.1\% \\
60+ & 54.8\% & 48.6\% & 51.4\% \\
\hline
\multicolumn{4}{l}{\textbf{Income Category} \textit{(N=4928)}} \\
High & 37.8\% & 30.8\% & 34.1\% \\
Medium & 22.4\% & 27.4\% & 25.0\% \\
Low & 35.0\% & 32.5\% & 33.7\% \\
Very low & 4.9\% & 9.2\% & 7.2\% \\
\hline
\multicolumn{4}{l}{\textbf{Psychosocial Factors (Prevalence)}} \\
PTSS (Positive) & 25.8\% & 42.7\% & 34.9\% \\
Anxiety (Positive) & 12.5\% & 21.5\% & 17.3\% \\
Depression (Positive) & 16.9\% & 18.3\% & 17.6\% \\
\hline
\end{tabular}
\end{table}
To reconstruct patient care trajectories, we aggregated and chronologically ordered all medical events for each patient, creating a sequence of labels. Events were categorized into three care settings: outpatient/ambulant (append label 'a' to this patient's sequence), hospitalization/inpatient ('h'), and day hospital/Tagesklinik ('t'). To account for temporal discontinuities in care, we introduced a 'break' event ('b') whenever the interval between consecutive encounters exceeded 21 days. Furthermore, medical encounters were enriched with clinical granularity by encoding events using ICD-10 codes corresponding to the 100 most frequent diagnoses in the cohort.

\subsection{Dynamic Time Warping and Clustering}

We employed Dynamic Time Warping (DTW) to quantify the dissimilarity between patient trajectories. Although originally developed for speech recognition~\cite{sakoeDynamicProgramming1978}, DTW has been successfully adapted for biomedical research to align heterogeneous clinical sequences that may vary in speed or temporal phase~\cite{giannoulaIdentifyingTemporal2018, allamAnalyzingPatient2021, forrestDynamicTime2025}.

To ensure biological plausibility and computational efficiency, we applied the Sakoe-Chiba band constraint~\cite{sakoeDynamicProgramming1978} with a radius of $r=3$. This global constraint restricts the warping path, ensuring that an event $i$ in sequence $A$ can only be aligned with an event $j$ in sequence $B$ if $|i - j| \leq 3$. This approach prevents the alignment of clinically unrelated events occurring at vastly different time points and reduces the computational complexity to $O(N)$~\cite{mullerDynamicTime2007}.

\begin{figure}[ht]
    \centering
    \includegraphics[width=0.8\textwidth]{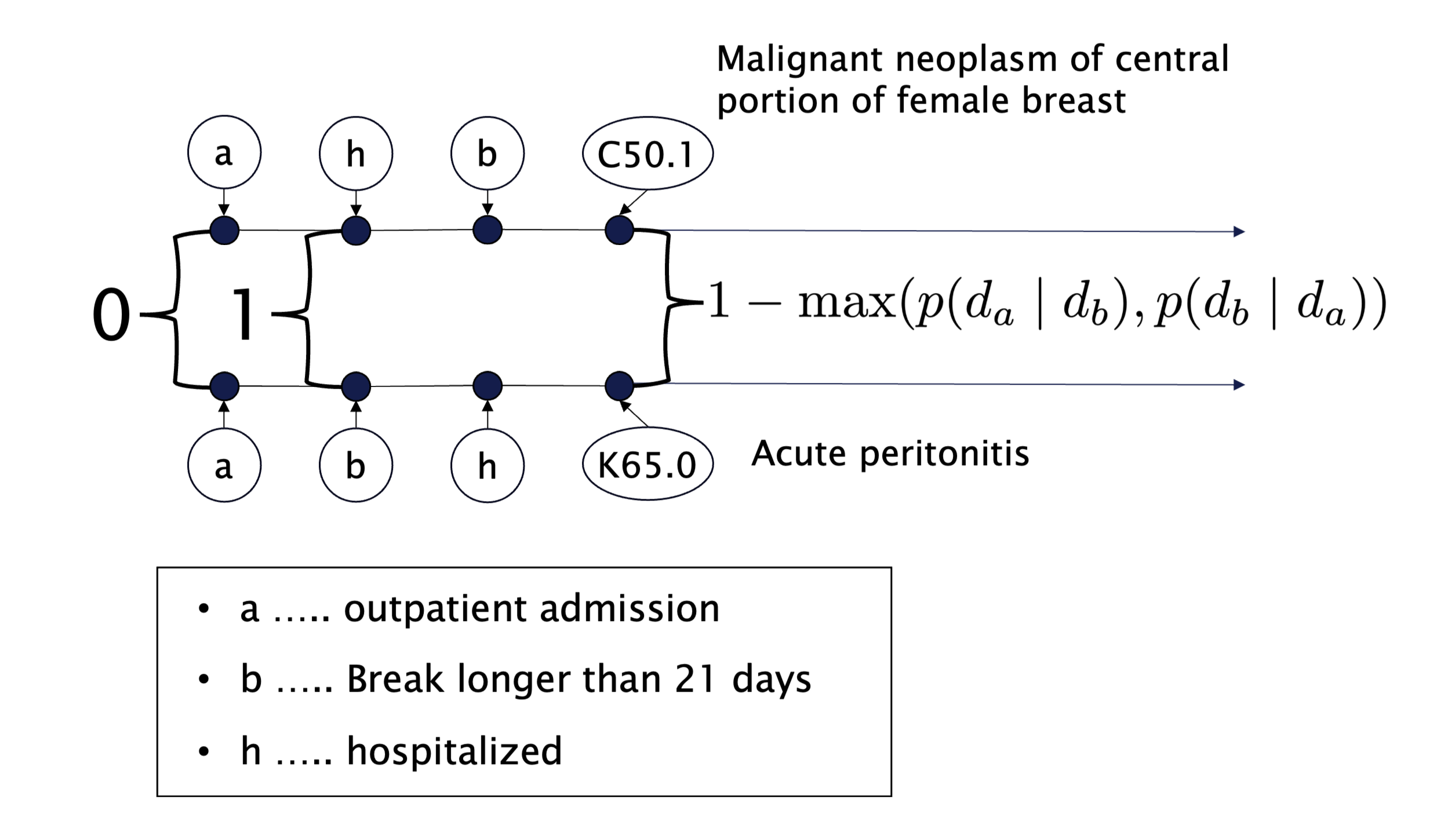}
    \caption{Conceptual diagram illustrating the dynamic time warping (DTW) distance metric. The cost calculation combines exact matches for event types (Kronecker delta behavior) with a probabilistic similarity measure for diagnoses, defined as $1 - \max(p(d_a \mid d_b), p(d_b \mid d_a))$. This enables the alignment of trajectories based on shared comorbidity patterns even when specific codes differ.}
    \label{fig:dtw_distance_metric}
\end{figure}

We defined a custom distance metric designed to capture both the structural type of care and the clinical similarity of diagnoses (Figure~\ref{fig:dtw_distance_metric}). The distance $d(e_a, e_b)$ between two events $e_a$ and $e_b$ is calculated as follows:

\begin{enumerate}
    \item \textbf{Identical Events:} If the events are identical (e.g., both are 'hospitalization' markers or the same ICD-10 code), the distance is 0.
    \item \textbf{Generic Mismatch:} If one or both events represent generic visit types (outpatient, inpatient, day clinic, or break) without specific diagnostic codes, and they are not identical, the distance is set to 1. This conservative penalty reflects the lack of shared clinical information.
    \item \textbf{Diagnostic Similarity:} When both events are encoded with specific diagnoses, the distance is derived from their co-occurrence probability:
    \[
       d(d_a, d_b) = 1 - \max\big(P(d_a \mid d_b),\, P(d_b \mid d_a)\big)
    \]
    Here, $P(d_a \mid d_b)$ represents the conditional probability of observing diagnosis $d_a$ given diagnosis $d_b$ within the dataset. This component allows the metric to recognize that distinct diagnoses (e.g., metastatic sites) may still be clinically related, assigning them a lower distance than unrelated conditions. Consequently, if one diagnosis invariably co-occurs with another, their calculated distance is 0.
\end{enumerate}

To mitigate the influence of outliers, we excluded the top and bottom 10\% of sequences based on trajectory length. We then computed the pairwise distance matrix for the remaining cohort using the constrained DTW metric. Patients were stratified into subgroups using agglomerative hierarchical clustering with Ward's linkage method, which minimizes within-cluster variance~\cite{giannoulaExploringLongterm2024, defossezTemporalRepresentation2014}. This process resulted in the identification of nine distinct patient clusters.

\subsection{Regression Analysis}

We performed multivariable regression analyses to evaluate the association between the identified trajectory clusters and key health outcomes: all-cause mortality, anxiety, and depression. Anxiety and depression were assessed at the start of treatment using standardized self-report instruments, and thus represent baseline mental health characteristics at the point of care entry. 

For binary outcomes (mortality), we utilized logistic regression to estimate odds ratios (OR) for clusters 2--9, with the largest cluster (Cluster 1) serving as the reference group. To control for confounding variables and isolate the specific effect of the care trajectory patterns, models were adjusted for socioeconomic factors (income, education, employment status)~\cite{bedirSocioeconomicDisparities2022, inglebyInvestigationCancer2022}, demographic variables (gender, age category), sequence length, and the presence of the 100 most common diagnoses. For continuous psychometric outcomes (anxiety and depression scores), we applied Generalized Linear Models (GLM) adjusting for the same covariates, as mental health burden is known to vary significantly across cancer populations~\cite{fangSelfratedHealth2024, zhaoSideEffects2022}.

\section{Results}

% --- 1. INTRODUCTORY PARAGRAPH ---
Our analysis identified nine distinct patient trajectories (Clusters 1--9). The demographic, clinical, and trajectory characteristics for each cluster are summarized in Table \ref{tab:cluster_characteristics}. Cluster~1 serves as the reference group (N=1,729), characterized by the shortest observation period (median 6.3 years), the lowest mortality rate (31.5\%), and moderate sequence lengths (median 91 events). In contrast, Clusters 2--9 show comparable and substantially longer observation periods (median range 10.9--15.6 years), reducing the potential for bias due to differential follow-up, but exhibit markedly different sequence lengths (median range 78--198 events) and consistently elevated mortality rates (44.2--60.3\%). This pattern suggests the existence of distinct care trajectories within the longer-observed patient population, which we explore below.

\begin{table}[H]
\centering
\caption{Demographic, Clinical, and Trajectory Characteristics by Cluster}
\label{tab:cluster_characteristics}
\resizebox{\textwidth}{!}{%
\begin{tabular}{lccccccccc}
\toprule
\textbf{Feature} & \textbf{C1} & \textbf{C2} & \textbf{C3} & \textbf{C4} & \textbf{C5} & \textbf{C6} & \textbf{C7} & \textbf{C8} & \textbf{C9} \\ 
\midrule
\textbf{N} & 1,729 & 1,099 & 1,323 & 186 & 179 & 360 & 853 & 215 & 578 \\ 
\textbf{Age, years} (mean $\pm$ SD) & 56.9 (14.9) & 60.9 (14.3) & 57.0 (15.2) & 62.7 (12.4) & 61.8 (11.6) & 62.2 (12.5) & 62.0 (14.1) & 58.6 (13.0) & 59.9 (14.0) \\ 
\textbf{Female} (\%) & 49.3\% & 56.0\% & 51.6\% & 49.5\% & 57.5\% & 57.2\% & 56.4\% & 55.8\% & 59.5\% \\ 
\textbf{Mortality} (\%) & 31.5\% & 49.3\% & 44.2\% & 58.6\% & 60.3\% & 59.2\% & 54.5\% & 51.2\% & 48.6\% \\ 
\textbf{Seq Length} (median [IQR]) & 91 [62-130] & 78 [53-110] & 103 [65-150] & 196 [164-227] & 198 [166-228] & 134 [104-168] & 78 [52-106] & 181 [146-210] & 139 [109-176] \\ 
\textbf{Obs Period, years} (median [IQR]) & 6.3 [4.6-9.0] & 11.4 [6.8-15.3] & 11.0 [6.8-14.1] & 15.6 [9.9-17.9] & 13.6 [8.1-17.8] & 12.3 [6.7-17.1] & 10.9 [6.2-15.7] & 13.2 [8.4-16.1] & 13.3 [7.7-16.6] \\ 
\textbf{Anxiety Score} (mean $\pm$ SD) & 6.3 (4.2) & 6.6 (4.5) & 5.9 (4.2) & 5.3 (3.6) & 5.2 (4.2) & 6.2 (4.5) & 6.3 (4.7) & 6.2 (4.1) & 5.8 (4.2) \\ 
\textbf{Depression Score} (mean $\pm$ SD) & 5.1 (4.2) & 5.8 (4.6) & 5.1 (4.5) & 5.4 (4.4) & 4.7 (4.2) & 5.9 (4.9) & 5.6 (4.5) & 5.3 (4.3) & 5.0 (4.2) \\ 
\bottomrule
\end{tabular}%
}
\end{table}
% --- 2. CLUSTER PROFILES (Integrated Narrative) ---
Based on these characteristics, the clusters represent distinct clinical phenotypes. Cluster~1 represents the baseline group, characterized by younger patients, the shortest observation period, and the lowest mortality rate. Cluster~2 consists of older patients (mean age 60.9) with moderate sequence lengths but the cohort's highest baseline anxiety and depression scores. Cluster~3 shares the demographic profile of the reference group but follows a longer trajectory, with extended observation periods ($>$11 years) and intermediate mortality.

Among the high-intensity profiles, Cluster~8 exhibits long healthcare sequences (median 181 events), similar to the highest-risk groups, but with somewhat lower mortality (51.2\%) than Clusters 4--7. Cluster~9, a predominantly female group (59.5\%) with long sequence lengths (median 139 events), likely captures stable, long-term maintenance therapy trajectories. The remaining groups (Clusters 4--7) represent the highest-risk phenotypes. Clusters 4 and 5, with the longest sequences in the cohort (median 196 and 198 events, respectively), suggest prolonged, complex care pathways. In contrast, Clusters 6 and 7 show shorter sequences (median 134 and 78 events) despite comparable observation periods, suggesting more rapid disease progression with fewer healthcare encounters. The distributions of sequence length and observation period across clusters are shown in Figure~\ref{fig:Violin}.

\begin{figure}[ht]
    \centering
    \includegraphics[width=0.9\textwidth]{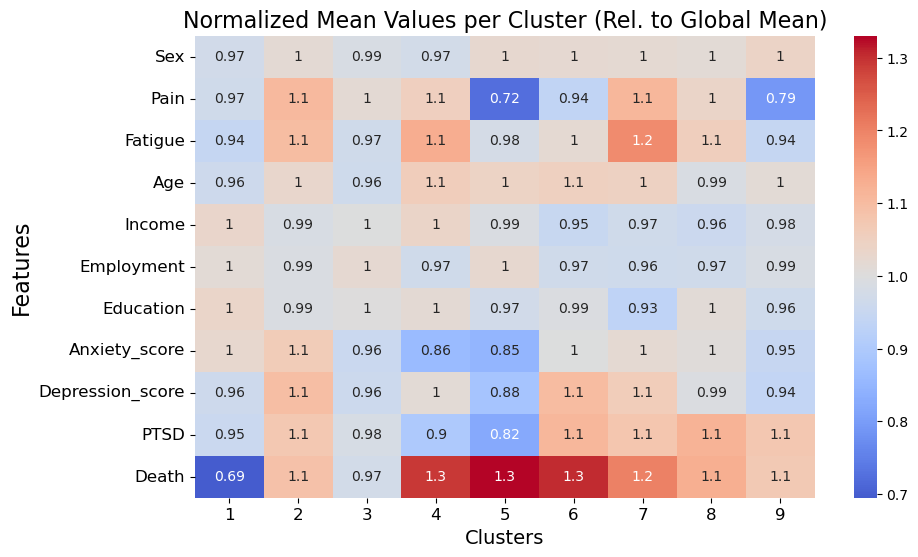}
    \caption{Heatmap displaying normalized mean values of various features across nine different clusters. The color scale ranges from blue (lower values) to red (higher values), highlighting significant deviations in feature values relative to the average of the remaining clusters.}
    \label{fig:ClusterDescription}
\end{figure}

% --- 3. PRIMARY FINDINGS (Mortality) ---
The primary finding of this study is that the identified clusters provide additional insights, particularly regarding mortality, beyond what is explained by the adjusted variables. As seen in the raw data (Table \ref{tab:cluster_characteristics}), Clusters 4, 5, 6, and 7 exhibit the highest unadjusted mortality rates (ranging from 54.5\% to 60.3\%). This trend persists in the regression analysis (Table \ref{tab:regression_summary_transposed}); these clusters exhibit significantly higher mortality odds compared to other clusters, as evidenced by the heatmap (Figure \ref{fig:ClusterDescription}) and the odds ratios (Table \ref{tab:regression_summary_transposed}). Notably, the odds ratios for mortality in these clusters show confidence intervals that do not overlap with an odds ratio of 1, underscoring statistically significant associations. This indicates that the clusters capture meaningful patterns related to mortality that are not fully accounted for by the adjusted covariates.
\begin{table}[H]
\centering
\caption{Multivariable Regression Results: Association of Clusters with Mortality, Anxiety, and Depression}
\label{tab:regression_summary_transposed}

\resizebox{\textwidth}{!}{% Shrinks table to fit the page width
\begin{tabular}{lcccccccc}
\toprule
\textbf{Outcome} & \textbf{C2} & \textbf{C3} & \textbf{C4} & \textbf{C5} & \textbf{C6} & \textbf{C7} & \textbf{C8} & \textbf{C9} \\
\midrule
\multicolumn{9}{l}{\textbf{Mortality}} \\
OR (95\% CI) & 1.68 (1.36-2.07) & 1.68 (1.38-2.05) & 3.38 (2.13-5.37) & 3.16 (1.98-5.05) & 2.43 (1.75-3.38) & 2.32 (1.82-2.97) & 1.82 (1.20-2.76) & 2.22 (1.70-2.89) \\
\textit{P-value} & $<$0.001 & $<$0.001 & $<$0.001 & $<$0.001 & $<$0.001 & $<$0.001 & 0.005 & $<$0.001 \\
\midrule
\multicolumn{9}{l}{\textbf{Anxiety}} \\
OR (95\% CI) & 1.03 (0.82-1.30) & 0.89 (0.72-1.12) & 0.65 (0.35-1.19) & 0.58 (0.31-1.08) & 1.09 (0.75-1.58) & 1.11 (0.84-1.46) & 0.52 (0.30-0.91) & 0.81 (0.59-1.11) \\
\textit{P-value} & 0.810 & 0.323 & 0.161 & 0.084 & 0.659 & 0.460 & 0.021 & 0.184 \\
\midrule
\multicolumn{9}{l}{\textbf{Depression}} \\
OR (95\% CI) & 1.42 (1.13-1.77) & 0.94 (0.75-1.17) & 1.08 (0.64-1.83) & 0.86 (0.49-1.50) & 1.49 (1.05-2.11) & 1.41 (1.08-1.84) & 0.96 (0.60-1.55) & 0.81 (0.58-1.11) \\
\textit{P-value} & 0.002 & 0.564 & 0.762 & 0.597 & 0.026 & 0.011 & 0.871 & 0.185 \\
\bottomrule
\end{tabular}%
}
\end{table}

% --- 4. SECONDARY FINDINGS (Mental Health & Utilization) ---
For mental health outcomes, several clusters also display statistically significant associations. Clusters 4 and 5, in particular, are characterized by lower mean HADS-Anxiety scores (5.3 and 5.2, respectively; Table~\ref{tab:cluster_characteristics}) and depression scores compared to the reference group (Cluster~1: mean HADS-A = 6.3), suggesting notable differences in mental health burden that align with specific cluster-based healthcare utilization patterns.

% --- 5. SOCIODEMOGRAPHICS & CONCLUSION ---
Sociodemographic characteristics vary only slightly across clusters (Figure \ref{fig:ClusterDescription}). The trajectory clusters remain significant predictors of mortality and baseline mental health even after adjusting for these covariates, indicating that the temporal structure of care pathways captures information independent of traditional risk factors.
% --- FIGURES ---

\begin{figure}[ht]
    \centering
    \includegraphics[width=0.9\textwidth]{seq_desc.png}
    \caption{Violin and box plots showing the distribution of sequence length (blue, left axis) and observation period in years (red, right axis) across 9 clusters.}
    \label{fig:Violin}
\end{figure}

\section{Discussion}

This study demonstrates that longitudinal healthcare utilization trajectories derived from routinely collected clinical data can identify clinically relevant patient phenotypes that capture distinct mortality risk profiles and baseline mental health characteristics. Using dynamic time warping--based clustering of more than 8,000 real-world cancer care trajectories, we identified nine distinct trajectory phenotypes that were associated with substantially different mortality rates and baseline mental health characteristics. Notably, these associations persisted after adjustment for demographic characteristics, socioeconomic indicators, diagnosis profiles, and overall sequence length, indicating that the temporal organization of care events contains prognostic information beyond that captured by traditional static variables. Together, these results suggest that trajectory-based representations provide an interpretable digital abstraction of disease course and care delivery, motivating a closer examination of the clinical patterns underlying the high-risk clusters identified in this study. 

The clinical interpretation of these high-risk clusters reveals two distinct patterns. Clusters 4 and 5 are characterized by longer sequences and extended observation periods, suggesting chronic disease courses with intensive healthcare utilization over time. These patients likely experienced progressive disease requiring frequent interventions, consistent with patterns seen in advanced or treatment-resistant cancers. In contrast, Clusters 6 and 7 showed high mortality despite shorter sequence lengths, indicating rapid disease progression or late-stage presentation with limited treatment window. This distinction has practical value: it separates patients with prolonged complex care needs from those with aggressive disease trajectories, allowing for more targeted resource allocation and earlier palliative care discussions where appropriate.

An unexpected finding was that Clusters 4 and 5, despite their elevated mortality risk, showed reduced anxiety and PTSD scores compared to other groups. Since these mental health assessments occurred at the beginning of treatment (see Methods), these lower scores reflect baseline characteristics rather than psychological adjustment over time. This counterintuitive pattern—low initial distress despite poor long-term prognosis—may stem from several factors. First, these patients are older on average, and older adults often demonstrate better baseline emotional regulation. Second, their specific medical presentations (such as the bone fractures common in Cluster 4) may have directed initial attention toward physical pain management rather than the immediate psychological shock often seen in rapidly worsening cases. Alternatively, these patients may simply have better coping skills or stronger social support networks from the outset. 

The diagnostic patterns across clusters provide additional clinical context. Cluster 4 showed a 14-fold higher prevalence of pathological fractures (ICD-10 code M84.4) compared to other clusters, strongly suggesting bone metastases and advanced disease burden. This aligns with the cluster's high mortality and extended healthcare utilization, as patients with skeletal complications typically require ongoing orthopedic, pain management, and oncological interventions. Other clusters also showed distinct diagnostic signatures (Figure \ref{fig:ClusterDiag}), though the relatively small variation in sociodemographic features across clusters (Figure \ref{fig:ClusterDescription}) reinforces that the trajectory patterns capture dimensions of disease progression and care complexity rather than simple demographic differences.

From a methodological perspective, this work extends previous applications of DTW in healthcare \cite{giannoulaIdentifyingTemporal2018, giannoulaExploringLongterm2024} by incorporating a custom distance metric that combines care setting types with diagnostic similarity based on co-occurrence patterns. This approach allowed us to recognize that different ICD codes can represent related clinical states (such as primary tumor and metastatic sites), producing more clinically meaningful clusters than purely structural sequence comparisons. The fact that these trajectory-based clusters remained significant predictors after adjusting for the diagnoses themselves suggests that the temporal ordering and spacing of events contains independent prognostic information, consistent with recent work showing that disease timing and treatment intervals influence outcomes \cite{franssenAssociationTreatment2021, kouTreatmentIntervals2023, weiImpactDiagnosistoTreatment2025}.

The clinical implications are substantial. Current risk stratification in oncology relies heavily on tumor characteristics, staging systems, and demographic variables \cite{meyerStratificationSystem2025, riviereProstateSpecificAntigen2024}. Our results suggest that adding trajectory-based clusters to these models could improve prognostic accuracy and help identify patients who need more intensive monitoring or early supportive care interventions. For instance, recognizing a patient's trajectory as similar to Cluster 6 or 7 early in their care could prompt earlier goals-of-care conversations and integration of palliative services. Similarly, identifying patients following Cluster 4 or 5 patterns might help allocate resources for long-term care coordination and address the specific complications associated with chronic, complex disease courses.

The integration of socioeconomic and psychosocial variables alongside temporal patterns addresses a gap in existing trajectory-based studies, which have often focused solely on clinical sequences \cite{houCardiacRisk2021, leeDevelopingMachine2022, masonPredictionMetastatic2021}. While we found that cluster membership provided prognostic information independent of income and education, these socioeconomic factors likely interact with trajectory patterns in complex ways. Socioeconomically disadvantaged patients may enter high-risk trajectories due to delayed diagnosis or treatment barriers \cite{gaigerCancerSurvival2022, redondo-sanchezSocioEconomicInequalities2022}, suggesting that interventions targeting social determinants of health could potentially shift patients toward lower-risk trajectory patterns. Future research should examine whether trajectory patterns mediate the relationship between socioeconomic status and cancer outcomes.

\subsection{Limitations}
Several limitations should be considered when interpreting these findings. First, converting medical records into event sequences necessarily loses information about exact temporal intervals between visits. While this simplification enables scalable comparison of heterogeneous longitudinal records and focuses the analysis on the structure and sequencing of care, more granular time-dependent modeling approaches might capture additional prognostic information. Similarly, restricting our analysis to the 100 most common diagnoses means that rare conditions or specific complications were treated as generic visits, potentially masking heterogeneity within clusters. This design choice represents a trade-off between clinical specificity and computational tractability.

The DTW algorithm's Sakoe-Chiba band constraint (radius of 3) was chosen to ensure computational efficiency and biological plausibility, but this window prevents matching events that occur far apart in the sequence. As a result, long-range temporal dependencies, such as associations between early treatment decisions and late adverse outcomes, may not be fully captured. More broadly, recent transformer-based approaches such as life2vec \cite{savcisensUsingSequences2024} have demonstrated that deep learning on event sequences can yield higher predictive accuracy for outcomes including mortality. However, these models operate largely as black boxes, whereas the DTW-based clustering used here produces interpretable trajectory phenotypes that can be directly inspected and linked to clinical characteristics.

The dataset spans 37 years (1985--2022), during which cancer screening, diagnostic techniques, and treatment protocols evolved substantially. Our analysis does not fully separate the effects of these temporal changes in medical practice from the natural progression patterns of disease. Consequently, some trajectory patterns may partially reflect era-specific care pathways rather than intrinsic disease dynamics alone. To assess the extent of this potential confound, we examined the calendar midpoint of each patient's observation window across clusters. Clusters 2--9 showed broadly overlapping interquartile ranges (approximately 2007--2014), indicating that these clusters are not artefacts of era-specific care practices (Supplementary Figure~\ref{fig:TemporalStability}). Cluster~1, whose interquartile range fell later (2016--2019), reflects its characteristically short follow-up rather than a distinct enrolment cohort. Similarly, some apparently short trajectories may result from administrative censoring (study end date) or loss to follow-up rather than completed treatment, making it difficult to distinguish true short-course patients from those with incomplete data.

We also lacked data on important psychosocial factors such as family support, coping strategies, and pre-illness mental health status, all of which are known to influence cancer outcomes \cite{sengerRoleCoping2024, songMediatingRole2025}. Although anxiety and depression were assessed at the start of treatment, no data were available on mental health before the onset of illness or on how these scores evolved over the course of care.

Finally, this study was conducted at a single institution in Vienna, Austria. The identified clusters may reflect local referral patterns, institutional care protocols, or regional healthcare system characteristics that limit generalizability. External validation across multiple institutions and healthcare systems will be necessary to determine whether these trajectory phenotypes are broadly reproducible or context-specific.

\section{Conclusions}

In summary, this study demonstrates that longitudinal healthcare utilization trajectories derived from routinely collected clinical data can be transformed into interpretable digital phenotypes that enhance risk stratification in oncology. By applying dynamic time warping–based clustering to real-world cancer care trajectories, we identified distinct patient subgroups with markedly different mortality and mental health profiles that were not fully explained by traditional demographic, socioeconomic, or diagnostic variables. These findings highlight the prognostic relevance of the temporal organization of care events and underscore the value of trajectory-based representations as a complementary layer to existing clinical risk models.

From a digital medicine perspective, the proposed framework offers a transparent and scalable approach for integrating longitudinal care patterns into precision oncology workflows. With further validation and refinement, trajectory-based stratification could support earlier identification of high-risk patients, inform care coordination and supportive interventions, and contribute to more personalized and equitable cancer care. Future work should focus on prospective validation, cross-institutional generalizability, and the development of clinically deployable tools that translate trajectory-derived insights into actionable decision support.

% \section*{Acknowledgments}

% \section*{Funding}

% Flush all pending floats before references
\FloatBarrier

% References section
\printbibliography

% \newpage
% \section*{Unresolved Comments}
% \listoftodos[Summary of Tasks]

\newpage
\appendix
\beginsupplement
\section{Supplementary Information}

\begin{figure}[ht]
    \centering
    \includegraphics[width=1.0\textwidth]{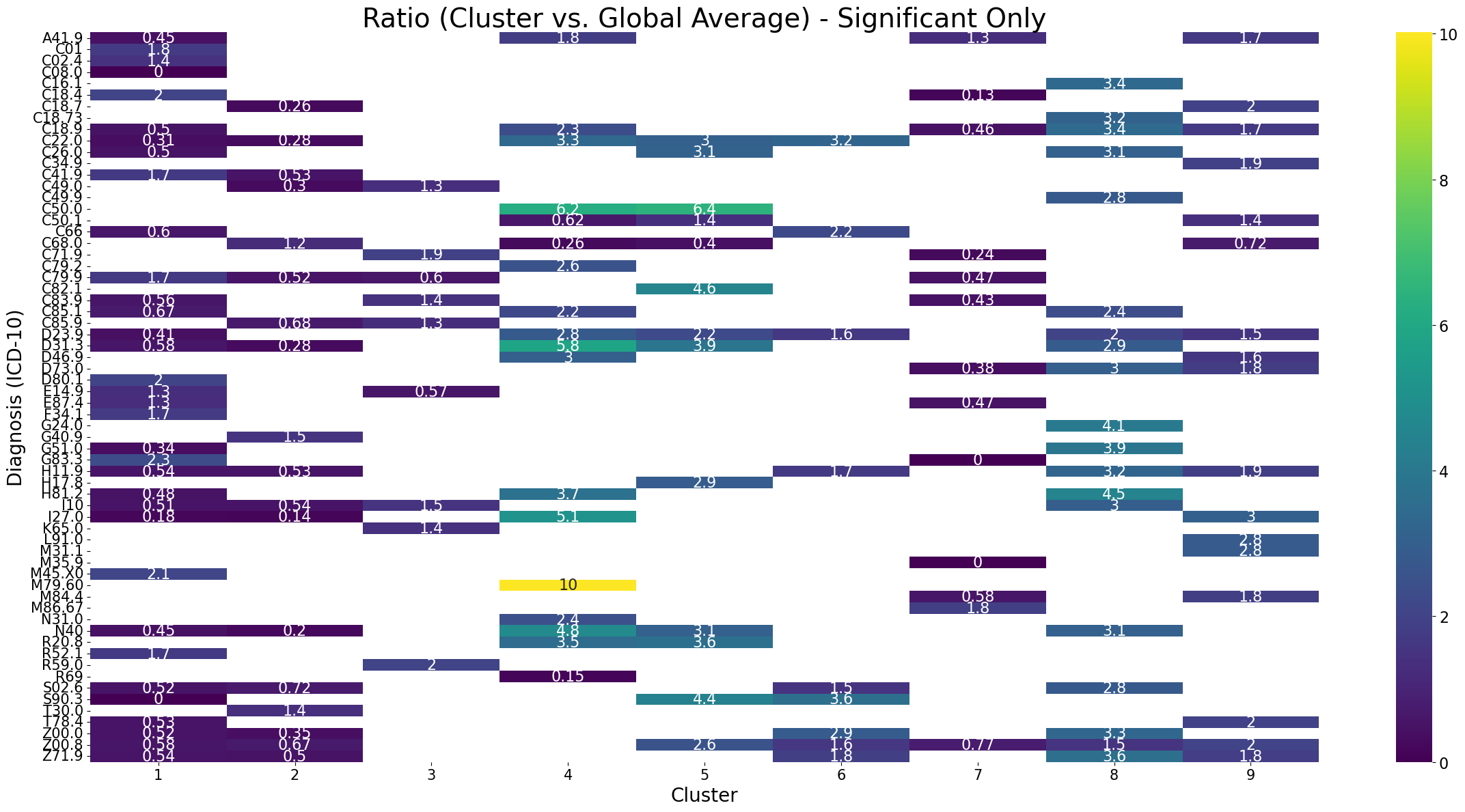}
    \caption{Heatmap displaying the prevalence of various diagnoses (ICD codes) across 9 clusters. Clusters 4 and 5 exhibit the most significant differences in diagnosis ratios; notably, pathological fractures (M84.4) occur 14 times more frequently in Cluster 4 compared to other clusters.}
    \label{fig:ClusterDiag}
\end{figure}

\begin{figure}[ht]
    \centering
    \includegraphics[width=0.9\textwidth]{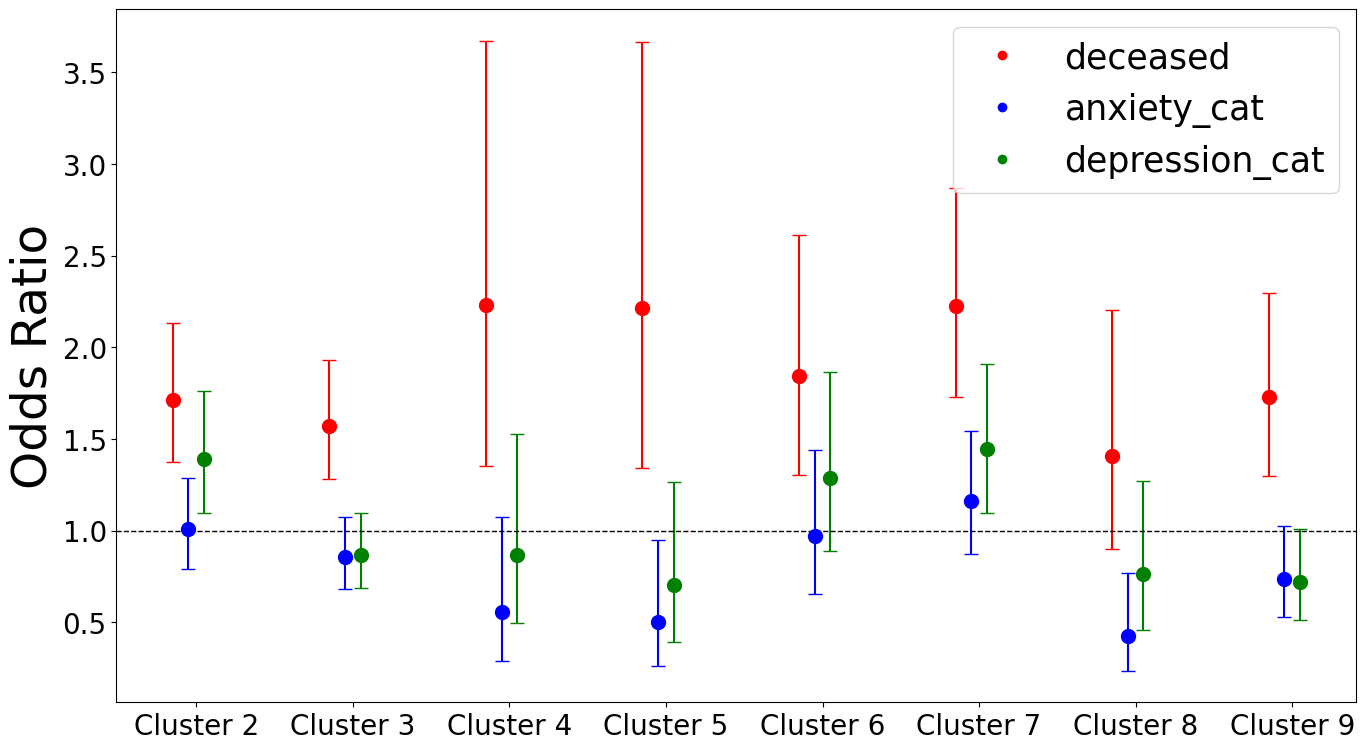}
    \caption{Odds ratios for deceased (red), anxiety (blue), and depression (green) across clusters 2--9 (Reference: Cluster 1). The model adjusts for income, education, employment, gender, age, sequence length, and top diagnosis. Clusters 4 and 5 show the highest mortality risk.}
    \label{fig:OR}
\end{figure}

\begin{figure}[ht]
    \centering
    \includegraphics[width=0.9\textwidth]{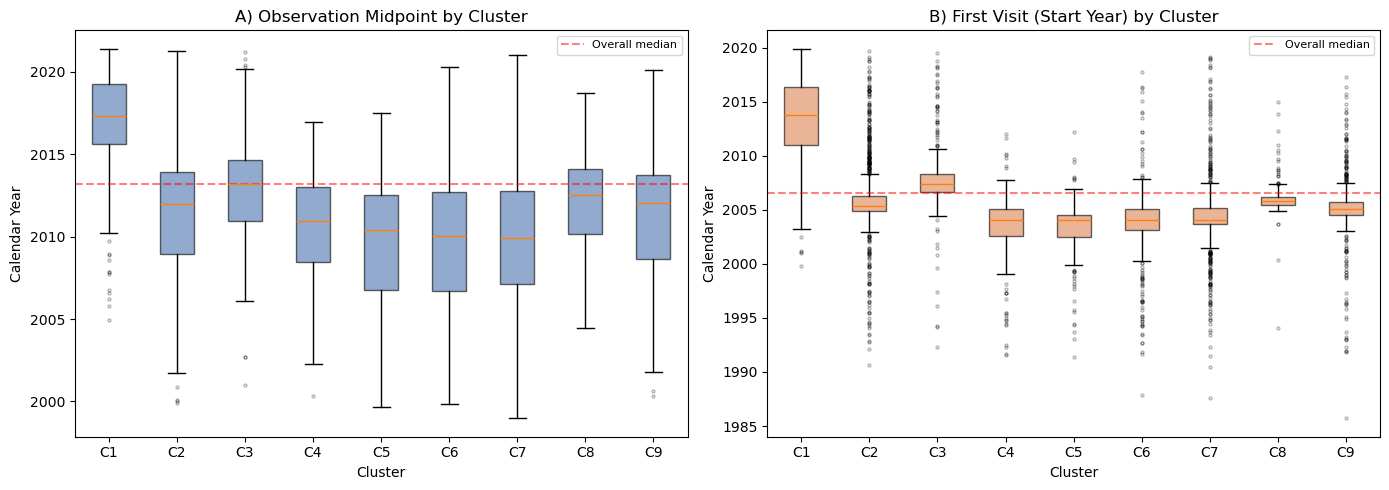}
    \caption{Temporal stability check. Box plots showing the distribution of (A) observation midpoint and (B) first visit year across clusters. Clusters 2--9 show broadly overlapping calendar periods, indicating that trajectory phenotypes are not artefacts of era-specific care practices. Cluster~1's later temporal profile reflects its characteristically short observation window.}
    \label{fig:TemporalStability}
\end{figure}

\end{document}